\documentclass[letterpaper, 10 pt, conference]{ieeeconf}  % Comment this line out if you need a4paper

\IEEEoverridecommandlockouts                              % This command is only needed if 
\usepackage{graphics} % for pdf, bitmapped graphics files
\usepackage{epsfig} % for postscript graphics files
\usepackage{mathptmx} % assumes new font selection scheme installed
\usepackage{times} % assumes new font selection scheme installed
\usepackage{amsmath} % assumes amsmath package installed
\usepackage{amssymb}  % assumes amsmath package installed
\usepackage{url}
\usepackage{array}
\usepackage{algorithmic}
\usepackage{algorithm}
\usepackage{subcaption}
\usepackage{cite}
\usepackage{bm}
\usepackage[table,xcdraw]{xcolor}

\newcommand{\PropMethod}{{DynEmph}}

\title{\LARGE \bf
  Dynamic Explanation Emphasis in Human-XAI Interaction \\ with Communication Robot
}

\author{Yosuke Fukuchi$^{1}$ and Seiji Yamada$^{1,2}$% <-this % stops a space
\thanks{$^{1}$Digital Content and Media Sciences Research Division, National Institute of Infomatics, Tokyo, Japan
        {\tt\small fukuchi@nii.ac.jp}.
$^{2}$The Graduate University for Advanced Studies, SOKENDAI, Tokyo, Japan.
This work was supported in part by JST, CREST, Japan, under Grant JPMJCR21D4.}%
}

\begin{document}

\maketitle
%\IEEEpeerreviewmaketitle

\thispagestyle{empty}
\pagestyle{empty}

\begin{abstract}
Communication robots have the potential to contribute to effective human-XAI interaction as an interface that goes beyond textual or graphical explanations.
One of their strengths is that they can use physical and vocal expressions to add detailed nuances to explanations.
However, it is not clear how a robot can apply such expressions, or in particular, 
how we can develop a strategy to adaptively use such expressions depending on the task and user in dynamic interactions.
To address this question, this paper proposes {\PropMethod},
a method for a communication robot to decide where to emphasize XAI-generated explanations with physical expressions.
It predicts the effect of emphasizing certain points on a user and aims to minimize the expected difference between predicted user decisions and AI-suggested ones.
{\PropMethod} features a strategy for deciding where to emphasize in a data-driven manner, relieving engineers from the need to manually design a strategy.
We further conducted experiments to investigate how emphasis selection strategies affect the performance of user decisions.
The results suggest that, while a naive strategy (emphasizing explanations for an AI's most probable class) does not necessarily work better,
{\PropMethod} effectively guides users to better decisions under the condition that the performance of the AI suggestion is high. 
\end{abstract}

\section{Introduction}
Explainable AI (XAI) technology has made great progress thanks to active trials in previous research.
XAIs can now generate a wide variety of explanations, including visual explanations using saliency maps~\cite{6875957},
 explanations about past decisions in similar conditions~\cite{VANDERWAA2021103404} or conversely, counterfactual ones~\cite{10.1145/3448016.3458455}.
Large language models (LLMs) can also generate a lot of natural language explanations that support AI predictions\cite{wiegreffe-etal-2022-reframing}.

This paper aims to discuss the problem of \textit{how} XAI-based decision support systems
utilizing a communication robot as a user interface should provide such explanations.
Communication robots have large potential in contributing to effective human-XAI interaction~\cite{MILLER20191,10160839}.
For example, they can enhance explanations with bodily and vocal expressions, adding detailed nuances that can enable richer and more effective communication.
However, poorly designed presentation strategies may lead to negative outcomes.
Explanation is a complex cognitive process~\cite{MILLER20191},
and it sometimes does not work or even result in undesirable results when influenced by the content of explanations, the status of the task, and the cognitive and psychological status of the users~\cite{maehigashi2023modeling,maehigashi2023roman,10.1145/3623809.3623834,ferguson2022explanations,herm2023impact}.

To address the problem of deciding how a social robot should provide XAI explanations, this paper proposes {\PropMethod},
a method for deciding where a communication robot should emphasize XAI-generated explanations with physical expressions.
The key concept of {\PropMethod} is that it decides where to emphasize with the aim of minimizing the expected difference
between a human decision and an AI-suggested one by referring to a user model that predicts how emphasizing would influence the human decision.
{\PropMethod} has two notable characteristics.
First, it can dynamically decide which point to emphasize by referring to the current and historical status of the task
and the interaction between a user and a social robot.
Second, {\PropMethod} features a strategy for deciding where to emphasize in a data-driven manner, 
so human designers do not need to manually design a strategy for selecting where to emphasize.
The design of influencing users through emphasis is inspired by libertarian paternalism~\cite{b35e72fa-fff9-37d3-a508-45875042aa96},
the idea of influencing one's choices to be better without coercing them.

This paper also reports the results of web-based user experiments 
that investigated human-XAI interaction with a communication robot that emphasizes explanations.
In a preliminary experiment, we compared three naive baselines.
With the ARGMAX strategy, a robot emphasizes explanations for an AI's most probable prediction class.
ROULETTE randomly chooses which explanations to emphasize using the probability of an AI prediction for each class.
FLAT does not emphasize explanations.
The results indicate an advantage of ROULETTE over ARGMAX in our setting, although ARGMAX is a common practice in XAI research~\cite{make3040048,rudin2019stop}.
Next, with a user model trained with the preliminary experiment's data, we implemented and evaluated {\PropMethod}.
To investigate both the maximum and practical potential of {\PropMethod},
we prepared two conditions: {\PropMethod}-ORACLE, in which an AI's suggested decisions are always correct,
and {\PropMethod}-RL, which actually uses reinforcement learning (RL) to infer AI-suggested decisions.
The results proved the concept of {\PropMethod}, that is, it successfully guides users to better decisions,
while we also found a potential risk of users evaluating the imperfection of {\PropMethod}-RL poorly despite of its suggestion performance being better than most of them.

\section{Background}
Presenting explanations along with AI predictions is beneficial for intelligent decision support systems (IDSSs)~\cite{10.1145/3610218,electronics12214430,10.1145/3581641.3584055}.
An increasing number of studies focus on user-centered approaches to XAI so that IDSSs can more effectively enhance user decision-making.
Typical studies investigate static aspects of interaction such as how different explanation types affect user performance~\cite{herm2023impact}, and fewer studies focus on problems with the dynamics of human-XAI interaction,
particularly the question of how to implement an XAI system that can adaptively change its behavior depending on the status of users and tasks.
SmartBBox~\cite{iros2023} attempts to dynamically communicate the reliability of autonomous driving by presenting fewer bounding boxes that indicate what the car is detecting in a continuous sequence of driving.
While SmartBBox considers binary judgments of whether a user rely on AI system,
{\PropMethod} targets concrete decisions of users who are getting a support of XAI-based IDSSs.

In addition, there is much room for discussion on the integration of social robots into human-XAI interaction~\cite{10160839}, in spite of its potential to achieve richer communication.
User-aware explanation between humans and robots is particularly studied in the research field of human-robot collaboration~\cite{10.5555/3171642.3171666}
because it is important to align beliefs and goals between humans and robots for successful collaboration.

The concept of {\PropMethod} is based on our previous trial~\cite{fukuchi2024dynamic},
in which X-Selector was proposed. It is a method for selecting which XAI explanations to present users in order to guide them to better decisions by predicting their decisions given specific combinations of explanations.
We aim to extend this approach from deciding what to explain to how to provide explanations in human-robot interaction.

\section{\PropMethod}
\subsection{Concept}
{\PropMethod} is a method for a social robot to dynamically decide which point of XAI-generated explanations to emphasize.
The concept of {\PropMethod} is that it aims to guide a user to an AI-suggested decision by predicting how emphasizing certain points will affect his/her decision. 

{\PropMethod} is formalized with two main components: UserModel and $\pi$.
UserModel is a computational model of a user who makes a decision $d_u$ given explanations with certain emphasis points.
\begin{equation}
  \mathrm{UserModel}(\bm{c}, \bm{x}, d) = P(d_u = d| \bm{c}, \bm{x}).
\end{equation}
UserModel outputs a probability distribution of $d_u$ conditioned by a set of explanations with or without emphasis ($\bm{x} = \{(x_i, e_i)\}$), and other contextual variables $\bm{c} = \{c_j\}$, where $x_i$ is a sample of explanation, and $e_i$ is a boolean that represents the decision of whether to emphasize $x_i$. 
%Let $e_i = 1$ when $x_i$ is to be emphasized and $0$ otherwise.
$\bm{c}$ includes other information such as AI predictions for decision support, task status, and user status. 
UserModel can be implemented with machine learning in a supervised manner.

$\pi$ is a model that infers an AI-suggested decision $d_\mathrm{AI}$.
The inference is done in parallel with user decision-making:
\begin{equation}
  \pi(\bm{c}, d) = P(d_\mathrm{AI} = d| \bm{c}).
\end{equation}
This paper assumes an RL model for implementing this.

{\PropMethod} aims to minimize the expected difference between the probability distributions of $d_u$ and $d_\mathrm{AI}$ by comparing possible $e$ candidates.
Explanations and the decisions of whether to emphasize them $\hat{\bm{x}}$ are calculated as:
\begin{equation}\label{eqn:distance} %%% Expected value?
   \hat{\bm{x}} = \mathrm{argmin}_{\bm{x}} E_d[|\mathrm{UserModel}(\bm{c}, \bm{x}, d) - \pi(\bm{c}, d)|].
\end{equation}
Calculating the argmin of this equation means that {\PropMethod} simulates how $e_i$ in $\bm{x}$ will change $d_u$ and explores the best one that guides $d_u$ to $d_{\mathrm{AI}}$ the most.
(Note that $\pi$ is independent of $\bm{x}$.)

\subsection{Implementation}
\begin{figure}[t]
  \begin{center}
    \begin{minipage}[b]{0.49\linewidth}
    \includegraphics[width=0.99\linewidth]{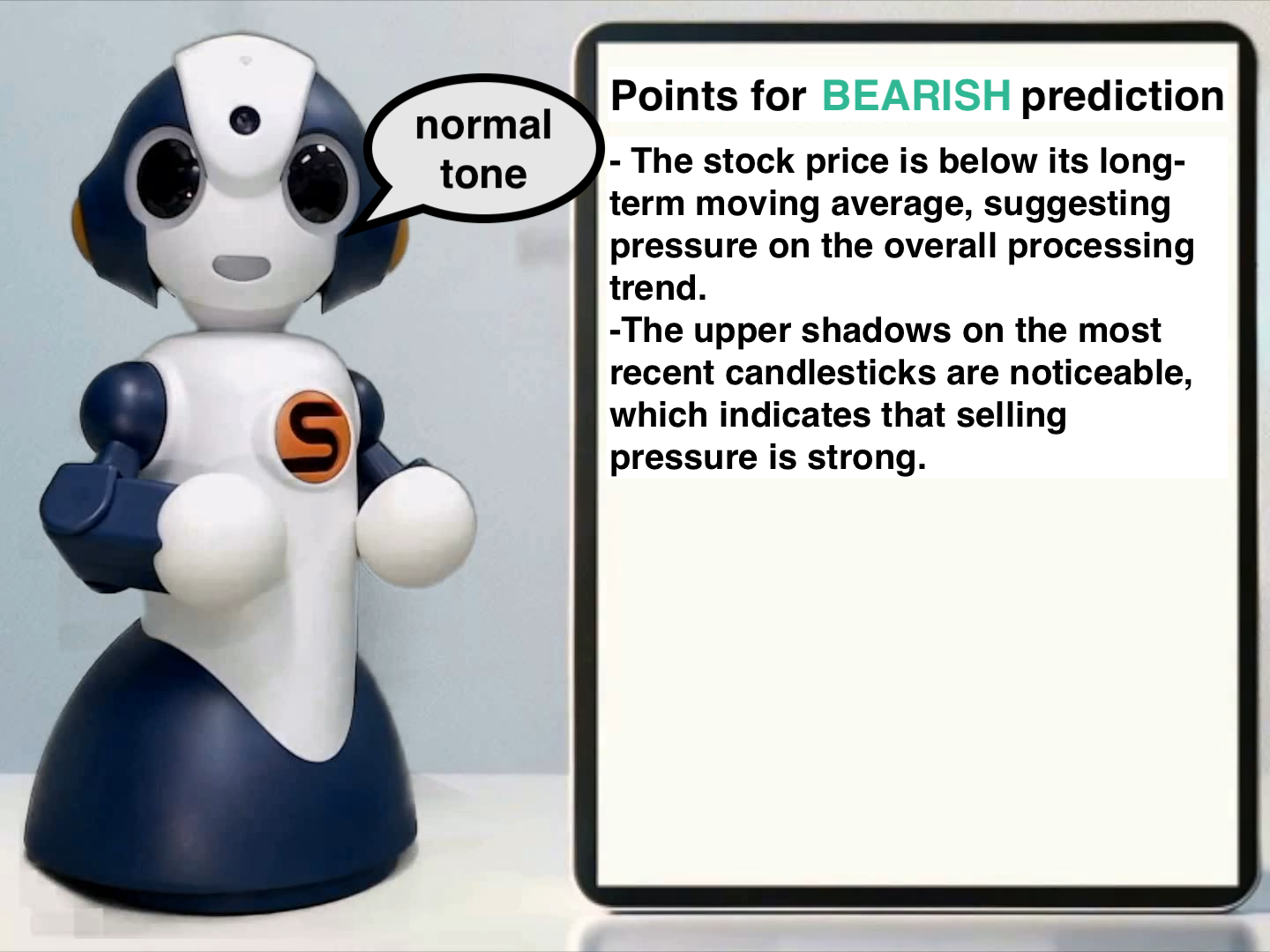}
    \label{fig:emphasis}
    \subcaption{Normal}
    \end{minipage}
    \begin{minipage}[b]{0.49\linewidth}
    \includegraphics[width=0.99\linewidth]{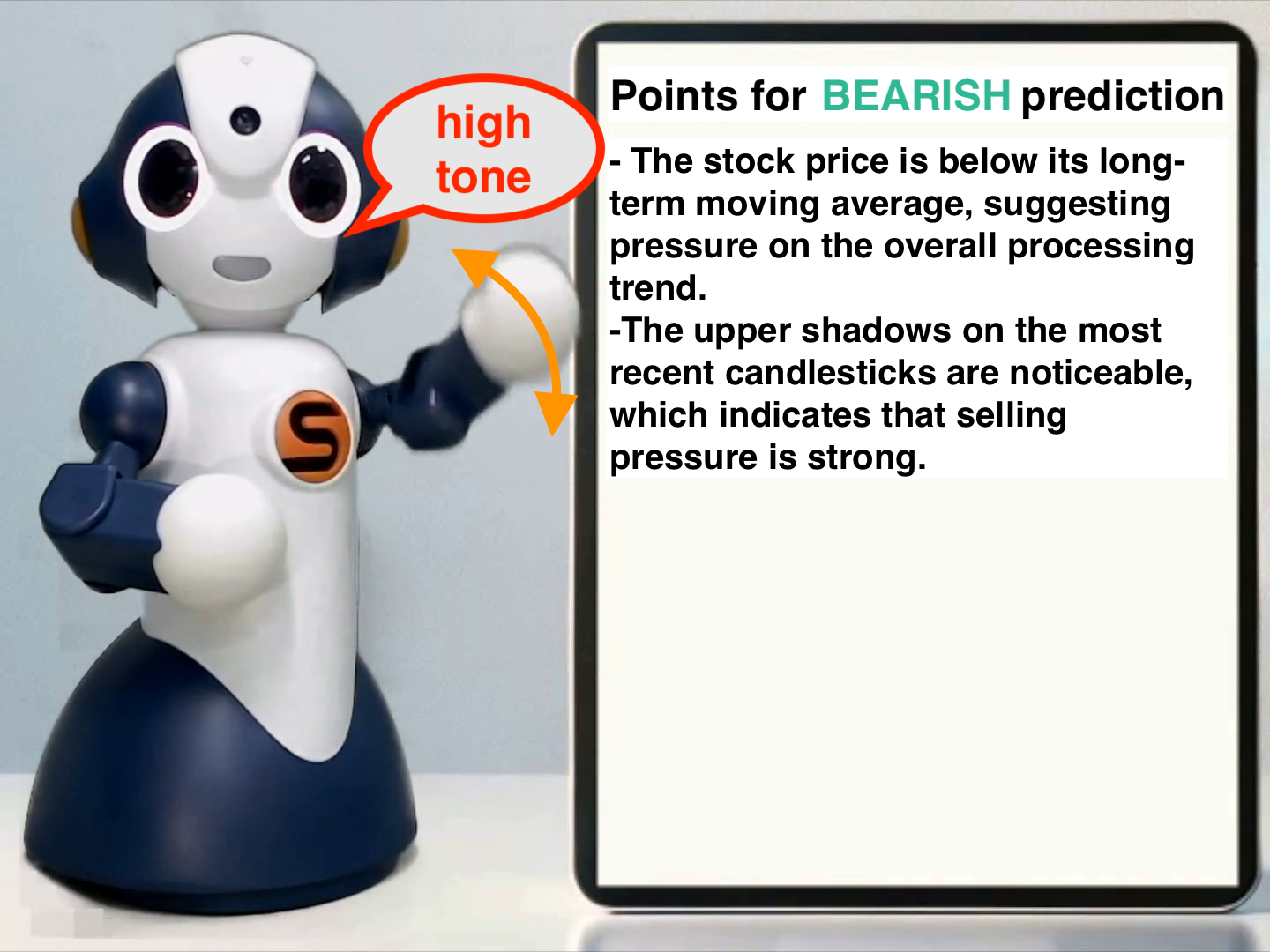}
    \label{fig:emphasis}
    \subcaption{Emphasis}
    \end{minipage}
  \caption{Robot emphasizes explanations about AI prediction with physical and vocal expressions}\label{fig:emphasis}
  \vspace{-6pt}
  \end{center}
\end{figure}
To implement and assess {\PropMethod}, we used an online stock trading simulator in which users get support from an XAI-based IDSS.
Each participant was allotted a virtual budget of one million JPY and engaged in stock trading over a 45-day period.
The trading interface provided a video in which a communication robot, Sota\footnote{https://sota.vstone.co.jp/home/}, introduces a stock price chart and an AI's prediction of future stock prices.
Then, it reads out the explanations of the reasons for the prediction.
In the video, an expression of emphasis by Sota was implemented with a higher voice tone, repetitive swinging of its left arm (about 2.8Hz), and lighting up of the LEDs on its eye contours (Figure \ref{fig:emphasis}).
We used VOICEVOX\footnote{https://voicevox.hiroshiba.jp/product/zundamon/}, a voice synthesizing software, because it has a feature to control voice tone by changing parameters.

Let $d_u$ be the amount of position that a user has after the day's order.
Trading could be done in units of 100 shares in accordance with Japan's general stock trading system.
%Because the participant's budget was around 1M JPY and the stock price was around 2K JPY, $d_u$ took the value of either 0, 100, ..., 500, and let $D = \{0, 100, ..., 500\}$ the set of possible values.
Given the participant's budget of approximately 1M JPY and an average stock price of 2K JPY, $d_u$ could take on values of 0, 100, ..., up to 500.
Thus, the set of possible $d_u$ values $D$ is $\{0, 100, ..., 500\}$.

The price prediction AI (StockAI) referred to an image of a candlestick chart and classified the prediction result of future stock prices into three classes: BULL (over +2\%), NEUTRAL (from -2 to +2\%), and BEAR (under -2\%)~\cite{7979885}.
Let $p$ be a vector that represents the calculated probabilities for each class.
It was implemented with ResNet-18~\cite{he2016deep} using the PyTorch library\footnote{https://pytorch.org/} and trained in a supervised manner.
For the training, we collected the data on historical stock prices from 2018/5/18 to 2023/5/16 of the companies that are included in the Japanese stock index Nikkei225.
The data were split by stock code, and we used three-quarters as the training dataset and the remainder as the test.
The model predicted the correct class among the three classes with an accuracy of 0.474,
and the accuracy for binary classification (the rate of matching the sign of the expected value of the prediction and that of actual fluctuations) was 0.63 for the test dataset.

We input the chart images to the GPT-4V model in the Open-AI API~\cite{openai2023gpt4} and acquired two free-text explanations that justified each prediction class (BULL, NEUTRAL, BEAR) for each chart.
{\PropMethod} decided whether to emphasize explanations for each class.
There were $2^3 = 8$ possible patterns, but we excluded the pattern in which all explanations were emphasized
because we anticipated that it would diminish the effect of the emphasis.
Therefore, {\PropMethod} explored seven patterns for each chart.

\begin{figure}[t]
  \begin{center}
    \begin{minipage}[b]{\linewidth}
      \center
      \includegraphics[width=0.999\linewidth]{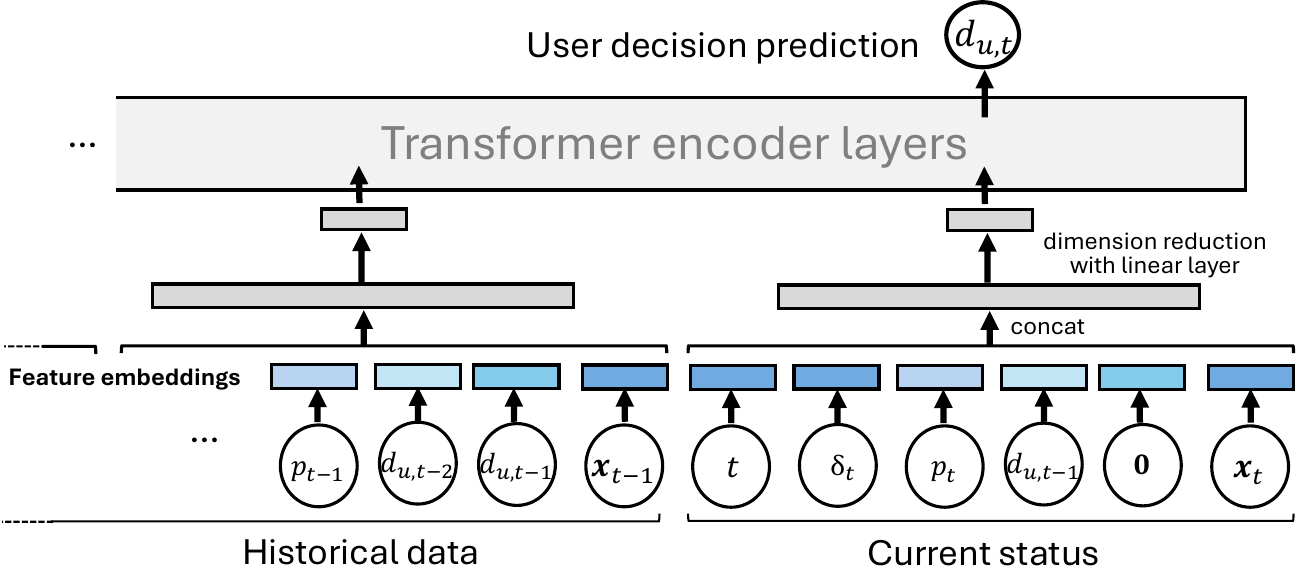}
    \subcaption{Structure of UserModel}\label{fig:model}
    \end{minipage}
    \begin{minipage}[b]{\linewidth}
      \center
      \vspace{6pt}
      \includegraphics[width=0.9\linewidth]{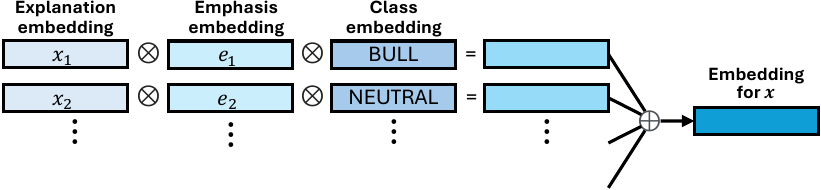}
    \subcaption{Calculation of embedding vector for $\bm{x}$. $\otimes$ and $\oplus$ represents element-wise product and sum, respectively.}\label{fig:x_embedding}
    \end{minipage}
    \caption{Prediction of $d_u$}
  \vspace{-6pt}
  \end{center}
\end{figure}
UserModel was implemented with a Transformer-encoder model~\cite{NIPS2017_3f5ee243} that can handle sequential information as an input.
UserModel referred to the historical and current information of ($\bm{c}, \bm{x}$).
Figure~\ref{fig:model} shows the structure.
Here, $\bm{c}$ is a tuple of variables that represent the date $t$, the percentage of the difference in total assets $\delta_t$, $p$, and the amount of stock before and after the day's order $d_{(u, t-1)}, d_{(u, t)}$.
$t$ and $d_u$ take discrete values, which were embedded with embedding layers in the PyTorch library,
whereas $\delta$ and $p$ are continuous and were transformed with linear layers.
The amount of stock after the current day's order was hidden.
By taking into account the collaboration history between a human and the target AI system, 
UserModel is expected to effectively capture the dynamics of decision support.
For encoding an explanation sentence $x_i$, we used the E5 (embeddings from bidirectional encoder representations) model~\cite{wang2022text} with pretrained parameters\footnote{https://huggingface.co/intfloat/multilingual-e5-large}.
The calculation of $\bm{x}$ is visualized in Figure~\ref{fig:x_embedding}. 
For each $i$, we calculated the element-wise product of the embedding vectors for $x_i$, $e_i$, and its prediction class.
The embedding for $\bm{x}$ is the element-wise sum of them.

$\pi$ for {\PropMethod}-RL was acquired with deep deterministic policy gradient (DDPG), a deep RL method~\cite{pmlr-v48-gu16}.
It infers $d_{\mathrm{AI}}$, that is, AI-suggested decision of how many position a user should have given $\bm{c}$.
Although DDPG outputs both mean and standard deviation values of an action, we used only mean for simplicity.
Note that unlike $d_u$, the value of $d_{\mathrm{AI}}$ is not necessarily in the discrete candidates of $D$ because DDPG can deal with continuous action space.

We calculated equation \ref{eqn:distance} with UserModel and $\pi$ above. 
Specifically, it calculates the expected absolute difference between $d_u$ and $d_{\mathrm{AI}}$:
\begin{equation}
  E[|\mathrm{UserModel}(...) - \pi(...)|] = \sum_{d \in D} P(d_u = d | \bm{c}, \bm{x}) \cdot |d - d_{\mathrm{AI}}|
\end{equation}

We trained $\pi$ to decide $d_{\mathrm{AI}}$, or how many position to have, to maximize assets on the basis of $p$s of the last three days with the training dataset.
The reward for the policy was calculated as the difference in total assets between the current day and the previous day.
The discount rate was set 0.6.
In this simulation, $d_{\mathrm{AI}}$ was not explicitly presented to users to more enhance the autonomy of their decision-making.
This design is inspired by libertarian paternalism~\cite{b35e72fa-fff9-37d3-a508-45875042aa96};
{\PropMethod} aims to nudge users to make better decisions while embracing the freedom of choice.
However, our formalization can be theoretically applied to a setting in which $d_{\mathrm{AI}}$ is given as well by including it with $\bm{c}$.

\section{Preliminary experiment}\label{section:preliminary}
\subsection{Overview}
We conducted a preliminary experiment to (i) collect data for training UserModel 
and (ii) investigated the trends of how the strategy of emphasis-point selection affected the user performance in the stock trading simulation.

We prepared three conditions:
(1) FLAT: providing explanations without emphasizing.
(2) ARGMAX: emphasizing the explanation for StockAI's most probable prediction class.
Targeting explanations about an AI's most probable prediction class is a common practice in XAI research, especially in classification tasks~\cite{make3040048,rudin2019stop}.
(3) ROULETTE: deciding whether to emphasize randomly depending on the probability of StockAI's prediction for each class.

\subsection{Procedure}
For the stock trading simulation, we selected a Japanese general trading company (code: 2768) from the test dataset
because of the common stock price range (1,000 - 3,000 JPY) and its high volatility compared with the other companies.
To control the accuracy of StockAI,
we calculated the moving average of the accuracy of StockAI with a window size of 45 and chose a section in which the accuracy of StockAI was 0.6 for the simulation.

We recruited participants to join the simulation with compensation of 200 JPY through Yahoo! Japan crowdsourcing\footnote{https://crowdsourcing.yahoo.co.jp/}
and got 225  participants.
The participants were first provided pertinent information, and 223 consented to the participation.
We gave them instructions on the task and gave basic explanations about stock charts and StockAI.
We instructed the participants to increase the given one million JPY as much as possible by trading with the IDSS's support.
We also motivated them by telling that the top performers would get additional rewards.
We asked a question to confirm that videos could be successfully played with audio and six questions to check their comprehension of the task,
and 168 passed the screening.
After familiarization with the user interface of the simulator, the participants traded for 45 virtual days successively.
152 participants completed the task (128 males, 32 females, and 2 did not answer; aged 20-79, $M = 47.1, SD = 10.7$).
We doubled the probability of assigning participants to the ROULETTE condition to use the data for training UserModel.
As a result, we got 36, 38, and 88 samples for FLAT, ARGMAX, and ROULETTE, respectively.
The procedure was approved by the ethics committee of National Institute of Informatics.

\subsection{Results}
\begin{figure}[t]
  \begin{center}
      \includegraphics[width=0.85\linewidth]{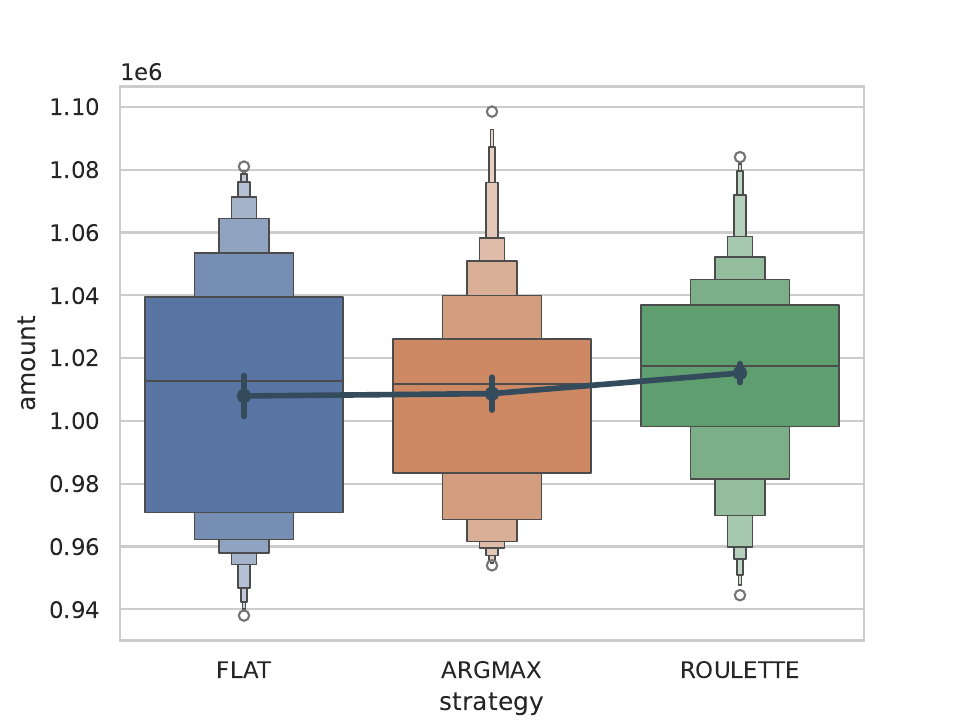}
    \caption{Final asset amount for each condition. Points and error bars represent means and standard errors.}\label{fig:preliminary_box}
  \end{center}
  \vspace{-8pt}
\end{figure}

Figure \ref{fig:preliminary_box} shows the final amounts of the participants' assets.
Here, we excluded one outlier whose final result was 1.221M JPY,
which was extraordinarily high compared with the results of both this and the next experiments, from ARGMAX.
The mean values of the FLAT, ARGMAX, and ROULETTE conditions were 1.0079M, 1.0087M, and 1.0152M in ascending order with standard deviations of 0.0392M, 0.0323M, and 0.0282M, respectively.

\begin{figure}[t]
  \begin{center}
      \includegraphics[width=0.85\linewidth]{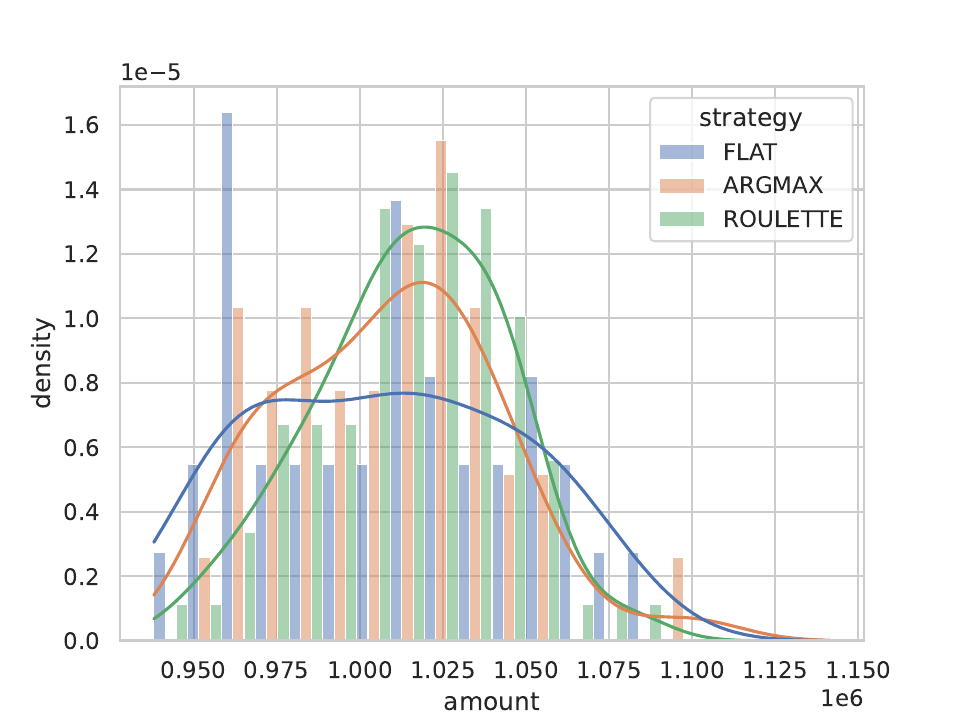}
    \caption{Histogram of distribution of final asset amount}\label{fig:preliminary_hist}
  \end{center}
  \vspace{-8pt}
\end{figure}

\begin{figure}[t]
  \begin{center}
      \includegraphics[width=0.7\linewidth]{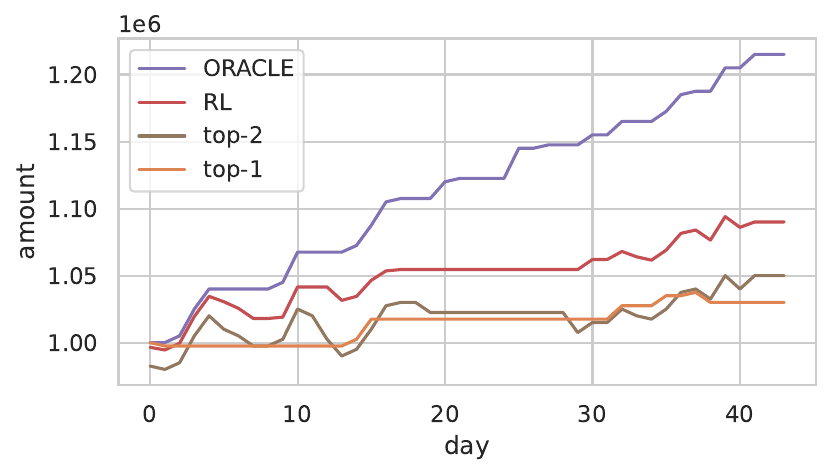}
    \caption{Performance of $\pi$ in ORACLE and RL conditions compared with naive strategies}\label{fig:performance}
  \end{center}
  \vspace{-8pt}
\end{figure}

Because it was difficult to statistically analyze the results due to the large variances,
we further investigated the distribution of the results.
Figure \ref{fig:preliminary_hist} illustrates the data.
We first found a wide variance for FLAT compared with the other conditions in which the robot emphasized explanations.
This result indicates that emphasizing has the effect of guiding users to specific decisions, making the result homogeneous.
ROULETTE was better than ARGMAX in terms of reducing the number of participants whose final assets were below 1M, that is, those who lost their assets in their trials. This resulted in ROULETTE's better mean score.
A possible reason of ARGMAX's underperformance is the complexity of $p$s.
Figure~\ref{fig:performance} the performance of $\pi$ (RL) and naive strategies in which one buys stocks when the most probable class of $p$ is BULL (top-1) and when it is either BULL or NEUTRAL (top-2).
Naive strategies failed to perform well though the better performance of RL, which also uses $p$s, suggests the informativeness of $p$s to get profit.
This highlights the difficulty of manually designing a strategy for deciding emphasis points depending on AI outputs and 
the necessity of developing a dynamic strategy.
We also need to note that, while emphasizing successfully raised the results of low-performance participants,
it also decreased the number of top-level performers through its guidance, which we will discuss in section \ref{section:future}.

\section{Evaluation of {\PropMethod}}
\subsection{Overview}
We further conducted a user study to evaluate {\PropMethod}.
We prepared two conditions: {\PropMethod}-ORACLE and {\PropMethod}-RL, and we compared the results with the ROULETTE baseline.
To prove our concept, {\PropMethod}-ORACLE simulated the optimal case in which $\pi$ suggests buying stocks if and only if the tomorrow's stock price goes up, excluding the influence of the performance of $\pi$.
%ORACLE is the upper bound of the profit possible in this term.
We also examined a more practical case with $\pi$ that was actually acquired with RL.
Figure \ref{fig:performance} shows the performance achieved with $\pi$s in the conditions.
The $\pi$ of {\PropMethod}-RL was able to earn 0.090M JPY, which is higher than most of the participants' results in the preliminary experiment.

\subsection{Procedure}
We conducted the experiment with the same procedure except for the strategy of emphasis-point decision.
153 participants were recruited.
107 passed the tests of video playing and task comprehension.
104 completed the task (79 males, 23 females, 2 did not answer; aged 20-75, $M = 48.9, SD = 12.1$).
51 and 53 participants were assigned to the {\PropMethod}-RL and {\PropMethod}-ORACLE conditions, respectively.

\subsection{Results}
\begin{figure}[t]
  \begin{center}
      \includegraphics[width=0.85\linewidth]{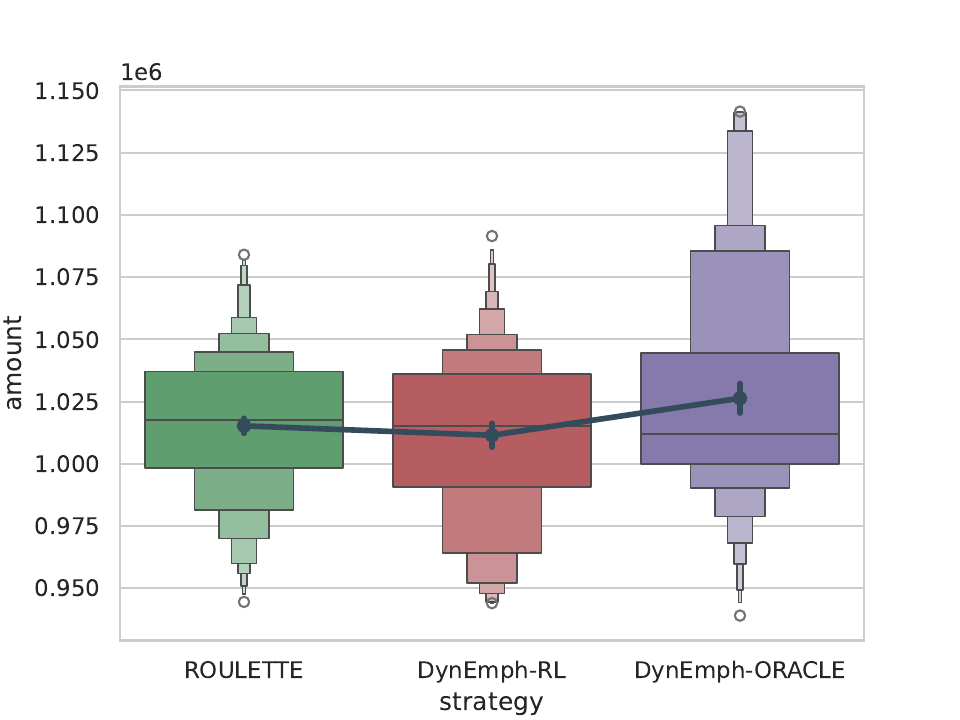}
    \caption{Final asset amount for each condition}\label{fig:main_box}
  \end{center}
  \vspace{-8pt}
\end{figure}

Figure \ref{fig:main_box} shows the results.
A conspicuous result is the outperformance of {\PropMethod}-ORACLE compared with ROULETTE. 
The mean and standard deviation were 1.0263M and 0.0434M, respectively.
{\PropMethod}-ORACLE largely raised the performance of top-25\% participants
and also improved the results of participants who lost their assets.
This proves {\PropMethod}'s concept that a social robot can guide users to better decisions by predicting the effect of emphasizing explanations and dynamically deciding where to emphasize.

However, {\PropMethod}-RL underperformed ROULETTE.
The mean and standard deviation were 1.0114M and 0.0345M, respectively.
Let us discuss this in the next section.

\section{Discussion and future work}\label{section:future}
\subsection{Investigating result of {\PropMethod}-RL}
Although the result of {\PropMethod}-ORACLE proved our concept, a challenge with {\PropMethod} that uses a more practical $\pi$ was also found in {\PropMethod}-RL's result.
To gain insights for future work, we conducted a further analysis of the results.

\begin{figure}[t]
  \begin{center}
      \includegraphics[width=0.85\linewidth]{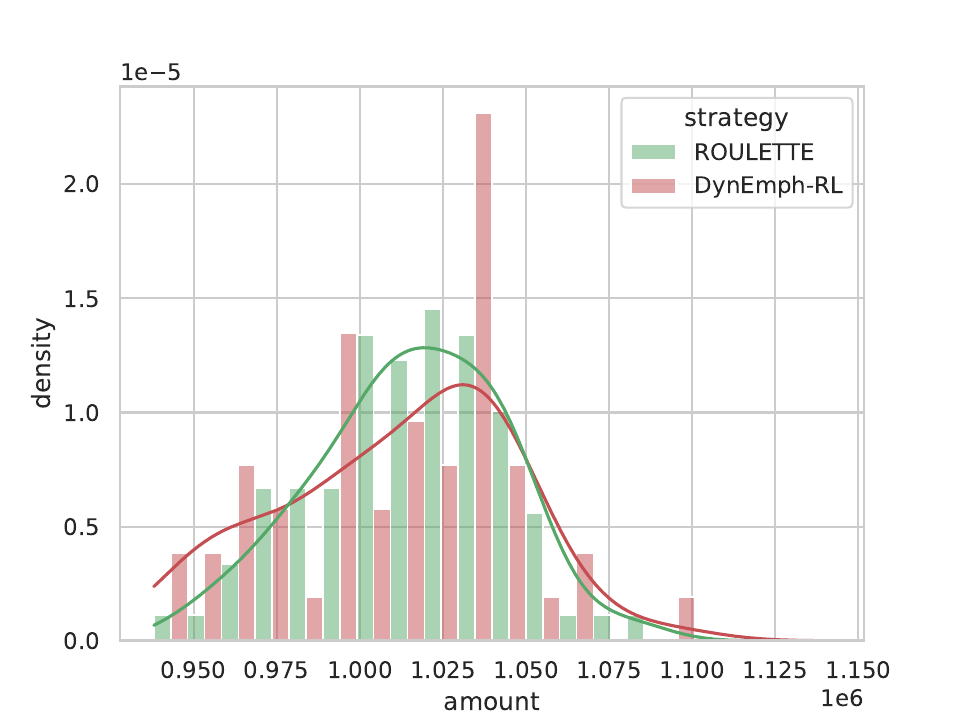}
    \caption{Histogram for comparison of ROULETTE and {\PropMethod}-RL}\label{fig:main_hist}
    \vspace{-6pt}
  \end{center}
\end{figure}
\begin{figure}[t]
  \begin{center}
      \includegraphics[width=0.75\linewidth]{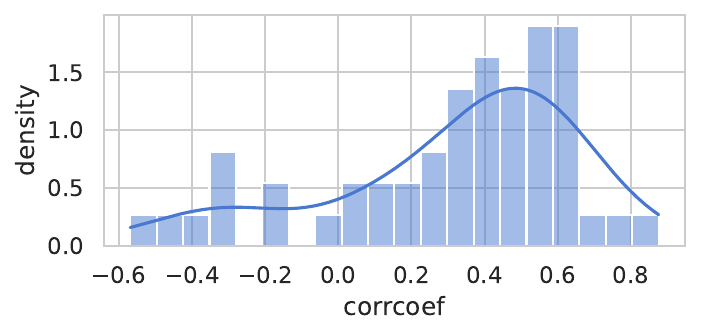}
    \caption{Correlation coefficient between $d_{\mathrm{AI}}$ and $d_{u}$ for each user}\label{fig:corrcoef}
    \vspace{-6pt}
  \end{center}
\end{figure}

Figure~\ref{fig:main_hist} shows the distribution of the final amounts in the {\PropMethod}-RL condition.
We found two noticeable tendencies.
First, {\PropMethod}-RL shifted the peak of the distribution to the right compared with ROULETTE,
indicating that {\PropMethod}-RL was effective for a portion of the population.
However, we also found an increase in the number of low-performance participants, which decreased the expected profit in this condition.
We also investigated the correlation coefficient between $d_{\mathrm{AI}}$ and $d_{u}$ for each participant.
This score is an indicator of whether {\PropMethod}-RL successfully guided users to $d_{\mathrm{AI}}$ through emphasis.
Figure~\ref{fig:corrcoef} visualizes the result.
While the values of most participants were positive, there was another cluster on the left,
meaning that some participants made decisions that were rather opposite the suggestions of the AI.
Because this score is strongly associated with the final asset amount ($r = .85, p < 10^{-14}$),
the participants in this cluster can be the next key target.

The performance of $\pi$ is presumably associated with this problem.
Although the $\pi$ of RL performed better than most of the participants, it was not perfect as we can see from the RL result in Figure~\ref{fig:performance}.
That is, participants might have thought that they were given false guidance to buy stocks during the task, which might have triggered distrust toward the robot. 
Unlike ROULETTE, whose nature of randomness makes the cues from the robot variable,
{\PropMethod} tended to consistently and repeatedly emphasize the same point when there was a large discrepancy between a user position and an AI suggestion,
which may have increased participants' feeling that the robot was intending to guide user decisions,
resulting in a more negative perception when guidance failed, and it may have strengthened the participants' attitude to avoid following the guidance.
ARGMAX also has a low-performance cluster (Figure \ref{fig:preliminary_hist}), which may be also caused for the same reason.

A possible solution to this is to adjust the strength of guidance.
{\PropMethod} in this paper always aims to guide user decisions toward AI-suggested ones,
but this can cause false guidance when the performance of AI is low.
Making guidance conservative depending on the confidence of an AI may work to avoid giving users the experience of mistaken guidance and instead earn their trust.
Combining {\PropMethod} with Trust-POMDP~\cite{10.1145/3359616}, with which a system can plan how to earn trust through AI actions, may be promising to proactively adjust the strength of guidance to maintain user trust.
After failure, trust repair~\cite{10.1145/3171221.3171258} including apologies and excuses may also be required for users who distrust the AI.

\subsection{Influence on high-performance users}
The results of section \ref{section:preliminary} also indicated the possibility that emphasizing explanations could make user performance homogeneous.
Though this is beneficial for users whose individual performances are low, it can be negative for top performers.
Adjusting the strength of guidance could be effective for this problem as well.
That is, {\PropMethod} may be able to avoid negative effects on top performers by assessing users' individual performance during a task and decreasing the strength of guidance from an AI.

\section{Conclusion}
To address the problem of how a social robot should provide XAI explanations, this paper proposed {\PropMethod},
a method for a robot to dynamically decide which points in XAI-generated explanations to emphasize with its physical expressions to influence users to make better decisions.
In {\PropMethod}, UserModel predicts how emphasizing certain points affects user decisions,
and {\PropMethod} aims to minimize the expected difference between predicted user decisions and AI-suggested ones inferred by $\pi$.
We conducted experiments to investigate how a strategy of emphasis-point selection affects the performance of user decision-making using a stock trade simulator with the support of an XAI-based intelligent decision support system.
As a result of a preliminary experiment, we found that a naive strategy of emphasizing explanations for an AI's most probable class does not necessarily work even though it is a common practice in human-XAI interaction.
Furthermore, an evaluation experiment proved the concept of {\PropMethod} that a robot can effectively guide users to better decisions by predicting how emphasizing certain points affects user decisions.
We also found a challenge with {\PropMethod} where users evaluate the imperfection of {\PropMethod}-RL poorly, which can cause user decision guidance to fail.
An insight from these results is that adjusting the strength of guidance through emphasis in explanations can be a promising way of avoiding both distrust and falsely guiding participants who can make better decisions solely by themselves.

\bibliography{bib}
\bibliographystyle{IEEEtran}

\end{document}